\documentstyle[prl,aps,epsf]{revtex}
\begin{document} \draft 
%\preprint{Presented in the XVI International Conference on Thermoelectrics, Dresden, Germany (1997).}
% The following line is crucial to get two-column format right
\twocolumn[\hsize\textwidth\columnwidth\hsize\csname @twocolumnfalse\endcsname
\title{Transport Coefficients of InSb in a Strong Magnetic Field}
\author{Hiroaki Nakamura, Kazuaki IKEDA$^a$, and
Satarou Yamaguchi}
\address{National Institute for Fusion Science (NIFS), Oroshi-Cho, Toki-City, Gifu-Prefecture, 509-52, Japan, Phone\&Fax : +81-52-789-4538, E-mail:hiroaki@rouge.nifs.ac.jp, http://rouge.nifs.ac.jp/\~hiroaki/index.html}
\address{$^a$Department of Fusion Science, The Graduate University for 
Advanced Studies (GUAS),
322-6, Oroshi-Cho, Toki-City, Gifu-Prefecture, 509-52, Japan}
\date{\mbox{This paper was presented in the XVI International Conference on Thermoelectrics, Dresden, Germany (1997).} } 
\maketitle

\begin{abstract}
Improvement of a superconducting magnet system makes induction of a strong magnetic field easier. This fact gives us a possibility of energy conversion by the Nernst effect. As the first step to study   the Nernst element, we measured the conductivity, the Hall coefficient, the thermoelectric power and the Nernst coefficient of the InSb, which is one of candidates of the Nernst elements. From this experiment, it is concluded that the Nernst coefficient is smaller than the theoretical values. On the other hand, the conductivity, the Hall coefficient ant the thermoelectric power has the values expected by the theory. 
\end{abstract}
%\pacs{75.10.Jm}
\vskip2pc] \narrowtext
\newcommand{\Tlow}{  T_{\rm low}  }
\newcommand{\Thigh}{  T_{\rm high}  }

\section{Introduction}
One of the authors, S. Y., proposed\cite{ref1,ref2} the direct electric energy conversion of the heat from plasma by the Nernst effect in a fusion reactor, where a strong magnetic field is used to confine a high temperature fusion plasma. He called\cite{ref1,ref2} the element which induces the electric field in the presence of temperature gradient and magnetic field, as  Nernst element. In his paper\cite{ref1,ref2}, he also estimated the figure of merit of the Nernst element in a semiconductor model. In his result\cite{ref1,ref2}, the Nernst element has high performance in low temperature region, that is, 300 - 500 K.
Before his works, the Nernst element was studied in the 1960's\cite{ref3}. In those day, induction of the magnetic field had a lot of loss of  energy. This is the reason why the Nernst element cannot be used. Nowadays an improvement on superconducting magnet gives us   higher efficiency of the induction of the strong magnetic field. We started  a measuring system of transport coefficients in the strong magnetic field to estimate efficiency of the Nernst element on a few years ago\cite{ref4}.
 As the first candidate of the Nernst element, we choose InSb, which is expected to have the high figure of merit according to the single-band model\cite{ref5}. The experiment results show that the Nernst coefficient is smaller than the theoretical values. On the other hand, the conductivity, the Hall coefficient and the thermoelectric power has the values expected by the theory.
In this paper, we introduce the experimental results and compare the theoretical calculations.
%%%%%%%%%%%%%%%%%%%%%%%%%%%%%%

\begin{table}
\begin{center}
\epsfxsize=4cm \epsffile{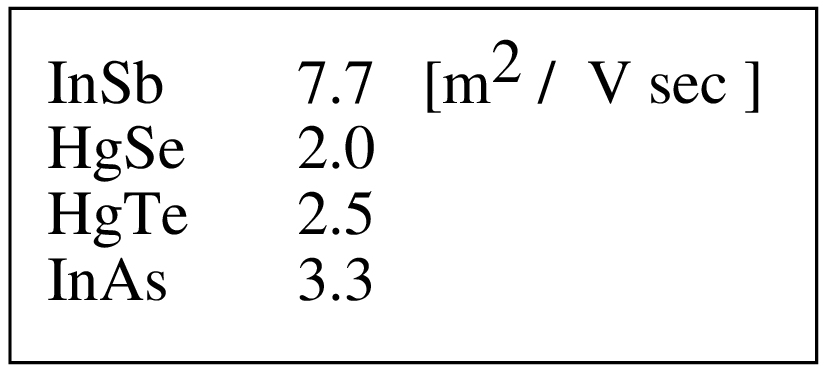}
%\epsfile{width=4cm,file=tab1.eps} 
\caption{Mobilities of electron near room temperature.}
\label{tab.1}
\end{center}
\end{table}
\newpage

%%%%%%%%%%%%%%%%%%%%%%%%%%%%%%
\begin{center}
\begin{figure}
\epsfxsize=6cm \epsffile{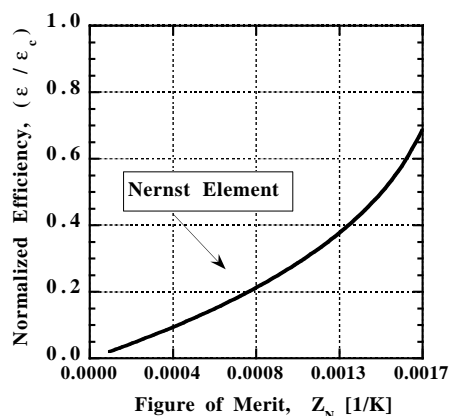}
%\epsfile{width=6cm,file=fig1.eps} 
\caption{The maximum efficiency of the Nernst element as a function of 
$Z_{\rm N}.$}
\label{fig.1}
\end{figure}
\end{center}

\section{Experiment}
\subsection{Choice of material}
We discuss the principle of transport phenomena in a magnetic field and a temperature gradient. This behavior is written by two phenomenological equations\cite{ref6} as follows:
\begin{eqnarray}
{\mbox {\boldmath $E$} } &=& \frac{ \mbox{\boldmath $J$} }{ \sigma}
                + \alpha \cdot \mbox{\boldmath $\nabla$ } T
 		+ R_{\rm H}  \mbox{\boldmath $B$}  
                   \times  \mbox{\boldmath $J$} ,  \label{eq.1} \\
{\mbox {\boldmath $q$} } &=& \alpha T \mbox{\boldmath $J$} 
                 - \kappa \mbox{\boldmath $\nabla$ } T
                 + N T  \mbox{\boldmath $B$}
                      \times  \mbox{\boldmath $J$}
                + L   \mbox{\boldmath $B$ } 
                    \times  \mbox{\boldmath $\nabla$} T, \label{eq.2}
\end{eqnarray}
where {\boldmath $E$} is electrical field, {\boldmath $J$} current density, 
{\boldmath $B$} magnetic field, $T$ temperature, $\sigma$  electrical conductivity,  
$\alpha$  thermoelectric power, $R_{\rm H}$ Hall coefficient,  
$N$  Nernst coefficient, $\kappa$  thermal conductivity and 
$L$ Righi-Leduc coefficient. The Nernst element uses the last term of eq.(\ref{eq.1}).

To simplify the discussion of the efficiency, we replace all transport coefficients by averaged quantities,
which do not depend on position within a device. 
In order to estimate efficiency of the Nernst element, 
it is useful to  define the  figure of merit $Z_{\rm N}$ 
as follows\cite{ref3}:
\begin{equation}
Z_{\rm N} \equiv \frac{\sigma B^2 N^2}{\kappa}. \label{eq.3}
\end{equation}
Using eq.(1), the optimal efficiency of thermomagnetic generators $\varepsilon_{\rm N}$
is given by\cite{ref3}
\begin{equation}
\varepsilon_{\rm N} = \varepsilon_{\rm C} 
    \left( 
       \frac{ 1-\delta^{\ast}_{\rm N} 
             }{ 1 +  \frac{ \Tlow }{ \Thigh } 
                \delta^{\ast}_{\rm N} } 
       \right), \label{eq.4}
\end{equation}
where   $\Thigh (\Tlow)$ is the temperatures of the heating (cooling) block, 
$\varepsilon_{\rm C}$ the carnot efficiency, $(\Thigh - \Tlow ) / \Thigh,$ and 
\begin{equation}
\delta^{\ast}_{\rm N} = 
   \sqrt{ 1 - Z_{\rm N} \left( \frac{\Tlow+\Thigh}{2}\right)}. \label{eq.5}
\end{equation}
The value of $\delta^{\ast}_{\rm N}$  must be a real number. 
This fact impose the following restriction as\cite{ref3}:
\begin{equation}
Z_{\rm N} \left( \frac{\Tlow+\Thigh}{2} \right)	 \leq 1.			\label{eq.6}
\end{equation}
We plot the normalized efficiency, $\varepsilon_{\rm N}/\varepsilon_{\rm C}$  
in Fig.~\ref{fig.1}
as a function of the figure of merit. This figure shows that 
$\varepsilon_{\rm N}$  increases monotonously as $Z_{\rm N}$ 
becomes larger. We, therefore, must choose the high-$Z_{\rm N}$  materials. We consider the transport coefficients to choose them. It is known from the Boltzmann equation that both conductivity and Nernst coefficient are proportional to Hall mobility\cite{ref7,ref8}. This fact derives the form\cite{ref1,ref4}:
\begin{equation}
Z_{\rm N} \propto \mu^3.
\label{eq.7}
\end{equation}
The equation (\ref{eq.7}) is a criterion for searching the Nernst element. 
Under this criterion, we first propose indium antimonide, InSb as a 
candidate of the Nernst element. To compare the mobility of InSb with 
the other materials, we summarize the values of the mobilities in 
Table~\ref{tab.1}.

%%%%%%%%%%%%%%%%%%%%%%%%
\begin{center}
\begin{figure}
\epsfxsize=8cm \epsffile{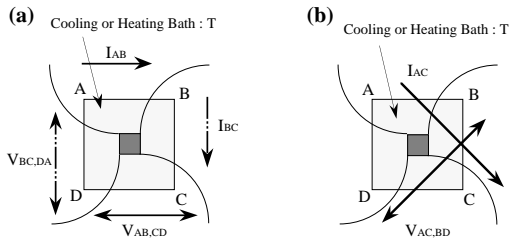}
%\epsfile{width=8cm,file=fig2.eps} 
\caption{Sample geometries for performing (a) resistivity and (b) Hall measurements by 
the van der Pauw method. Size of sample is 4mm $\times$ 4mm $\times$ 1mm.}
\label{fig.2}
\end{figure}
\end{center}
\newpage
\begin{center}
\begin{figure}
\epsfxsize=6cm \epsffile{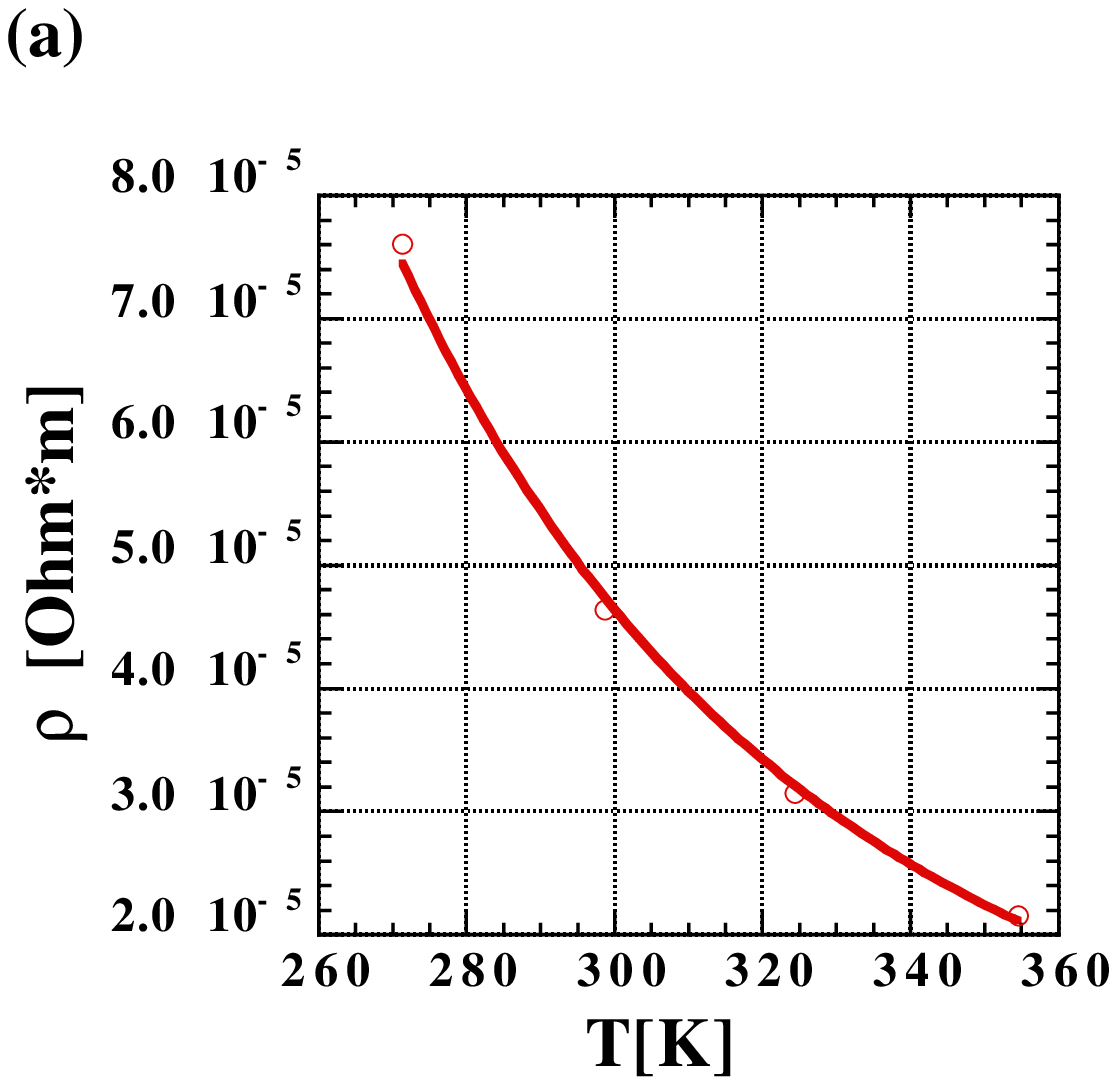}
\epsfxsize=6cm \epsffile{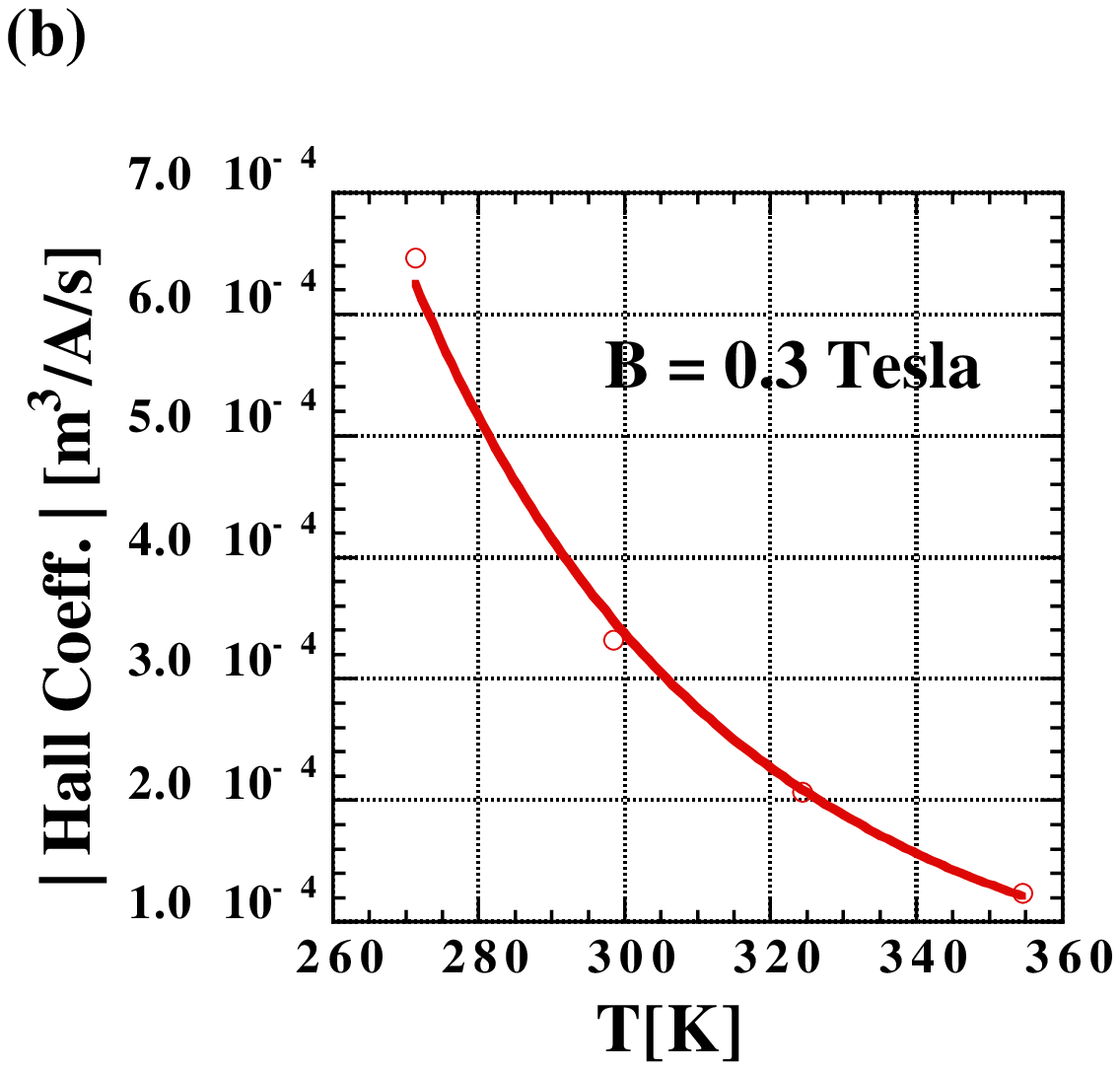}
%\epsfile{width=6cm,file=fig3a.eps} 
%\epsfile{width=6cm,file=fig3b.eps} 
\caption{Resistivity (a) and Hall coefficient (b) of InSb sample
as a function of temperature. The closed circles indicate the experimental 
results. Ths solid curve is a guide of eyes.}
\label{fig.3}
\end{figure}
\end{center}

%%%%%%%%%%%%%%%%%%%%%%%%%%%%%%
\subsection{Measurement of transport coefficients and results}
\subsubsection{Conductivity and Hall coefficient}
The carrier concentration of the InSb crystals investigated is $6.6 \times 10^{20}$~[ m$^{-3}$]
and its mobility 21 [m$^{2}$/V/s]  at 77K. 
This sample exhibited intrinsic behavior near room temperature. 
Copper wires with 50 $\mu$m-diameter are spark-bonded onto a crystal by using a capacitor discharge. Chromel-Alumel thermocouples, 
0.5 mm in diameter, are contact to heating 
and cooling units with  silver epoxy.
The temperature of the sample is controlled within 270-370K by the heat bath,
the water temperature  of which is kept a constant. 
We induced a strong magnetic field up to 4 Tesla by the superconducting coil 
to measure magnetoresistance of the sample. 
Analog signals of the thermocouples and 
voltage source are amplified and converted to digital data. 
The personal computer acquires these data and draws figures in real time.
We use the van der Pauw method\cite{ref9} 
to measure the conductivity and Hall coefficients. 
A geometry for the van der Pauw method is shown in Fig.~\ref{fig.2}.
Figure Fig.~\ref{fig.3}(a) shows the temperature dependence 
of the resistivity at B=0 Tesla. 
The temperature dependence of the Hall coefficient is represented in Fig.~\ref{fig.3}(b). 

\begin{center}
\begin{figure}
\epsfxsize=6cm \epsffile{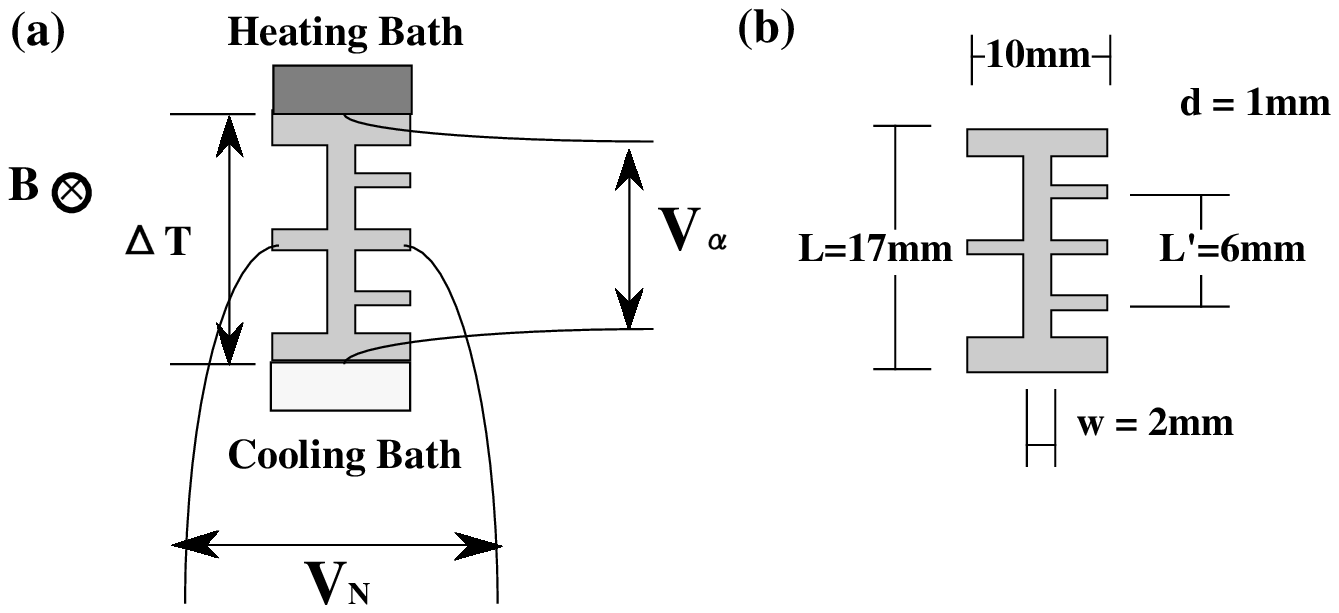}
%\epsfile{width=6cm,file=fig4.eps} 
\caption{Shape of the sample for measuring the thermoelectric power and the Nernst effect.
This shape is called the ``bridged shape".}
\label{fig.4}
\end{figure}
\end{center}
%%%%%%%%%%
\begin{center}
\begin{figure}
\epsfxsize=6cm \epsffile{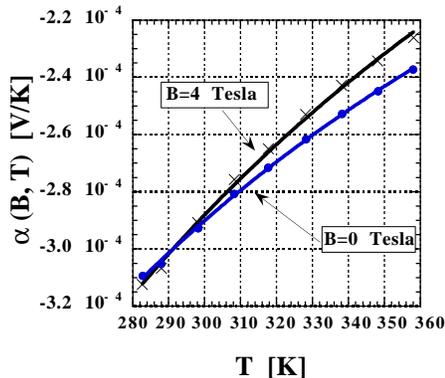}
%\epsfile{width=6cm,file=fig5.eps} 
\caption{Thermoelectric power of InSb sample as a function of 
temperature. The crosses represent experimental results at $B=0$ Tesla. 
The filled circles indicate the experimental results at $B=4$ Tesla. 
The solid curves are guide for eyes.}
\label{fig.5}
\end{figure}
\end{center}

\subsubsection{Thermoelectric power and Nernst coefficient}
The sample of measurement of  the thermoelectric power and the Nernst element 
is the same material as the van der Pauw method. 
However, the shape of the sample is changed from the square 
to the bridged shape  (Fig.~\ref{fig.4}).

\begin{center}
\begin{figure}
\epsfxsize=6cm \epsffile{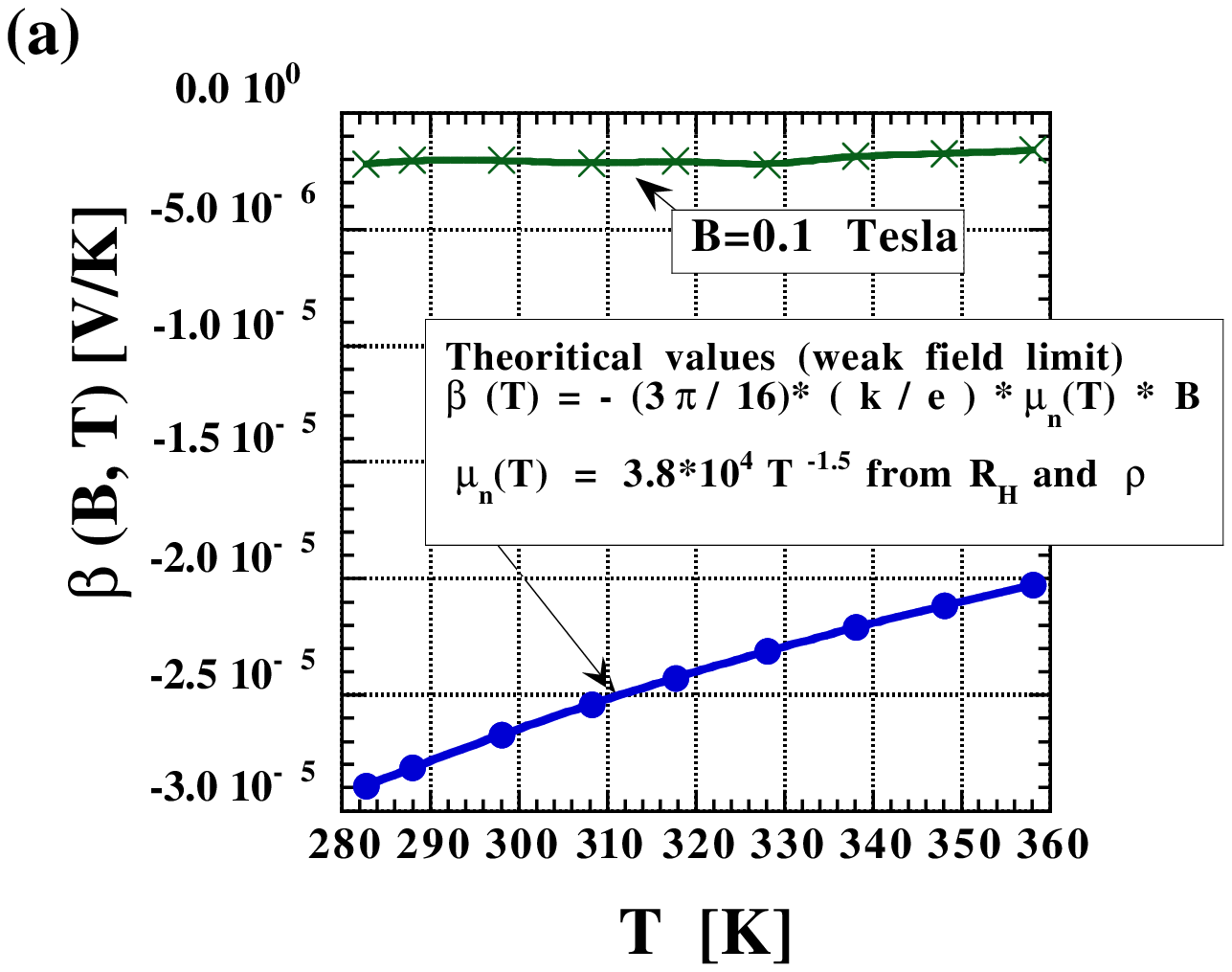}
\epsfxsize=6cm \epsffile{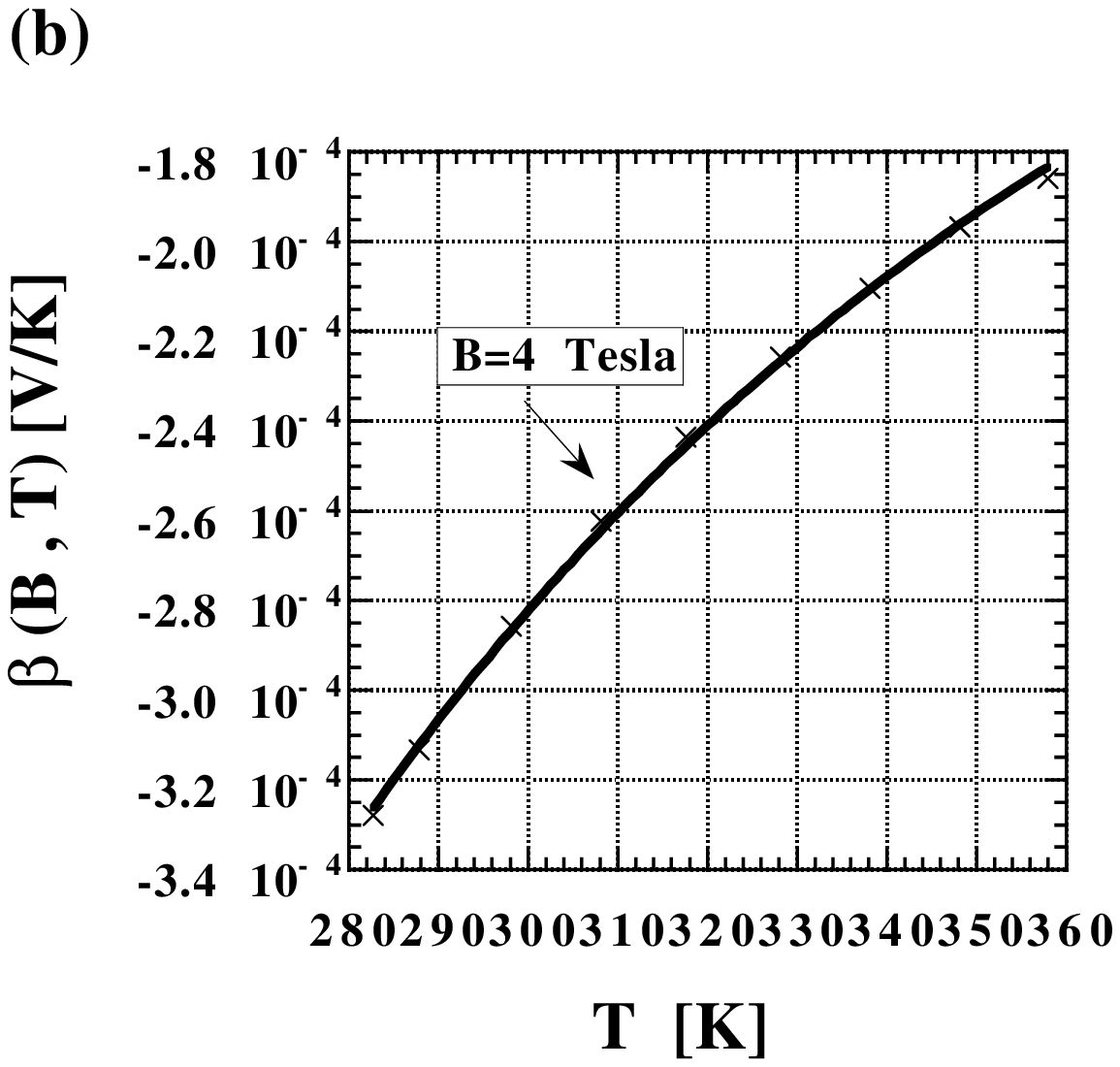}
%\epsfile{width=6cm,file=fig6a.eps} 
%\epsfile{width=6cm,file=fig6b.eps} 
\caption{Plot of Nernst effects, $\beta=N\times B$
in the case of $B=0.1$ Tesla (a) and 4 Tesla (b).
The crosses indicate the experimental results.
The filled circles in Fig.~\ref{fig.6}(a) were calculated by the single
band model with the mobilities which were given from the Hall coefficients and
resistivities. In Fig.~\ref{fig.6}(b), the solid curve is given by the least square 
method.}
\label{fig.6}
\end{figure}
\end{center}

In order to make temperature gradient in the sample, 
we used thermofoil heater for a heating copper block side, 
the water temperature of which is controlled by a low temperature incubator,
for a cooling one. Using the heating and cooling units,
the thermal difference 
across the sample was within 10-100K.
The thermoelectric  voltage, $V_{\rm \alpha}$ and the Nernst one, 
$V_{\rm N}$ have the following relations between the thermoelectric power 
and the temperature gradient as
\begin{eqnarray}
V_{\rm \alpha} &=& L \alpha | \mbox{\boldmath $\nabla $ } T | 
              \approx \alpha \Delta T,			\label{eq.8} \\
V_{\rm N} &=& w N B | \mbox{\boldmath $\nabla $ } T | 
              \approx \frac{w N B \Delta T }{L},	\label{eq.9}  
\end{eqnarray}
where $\Delta T$ is $(\Thigh - \Tlow ),$  $w$ the width of the sample 
and $L$ the length defined by Fig.~\ref{fig.4}. 
Here we define the following physical quantity, $\beta$ to compare 
the thermoelectric power and the Nernst effect:
\begin{equation}
\beta \equiv NB,	\label{eq.10}
\end{equation}
which has the same dimension, [V/K]  as $\alpha.$
The results of the measurement of 
$\alpha$ and $\beta$ in Figs.~\ref{fig.5}--\ref{fig.6}.
The thermoelectric power doesn't change very much as the magnetic field is induced. 
On the other hand, the $\beta$ depends on the magnetic field very much. 
In Fig.~\ref{fig.6}(a), we plot the results as the crosses and the theoretical 
values as the filled circles. The theoretical values are explained in the later.
The difference between the experimental results and the theoretical ones 
is the order  of 10. For the strong magnetic field, 
the results are shown in Fig.~\ref{fig.6}(b). 

\section{Analysis and physical quantities}
\subsection{Carrier concentration}
\begin{center}
\begin{figure}
\epsfxsize=7cm \epsffile{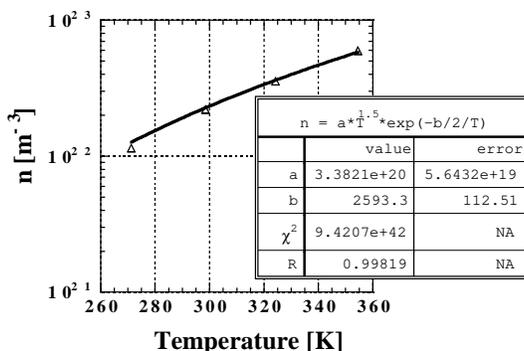}
%\epsfile{width=7cm,file=fig7.eps} 
\caption{\fussy Temperature~dependencd~of~the carrier concentration of~the~electron\@. 
{ \fussy Fitting~function~is 
$ n(T) = 3.3 \times 10^{20} T^{1.5} \exp \left( - \frac{2600}{k T}\right). $  } 
Comparing the intrinsic concentration eq.~(\ref{eq.13}),
we have $ E_{\rm G} \approx 0.22 \mbox{ [eV] } $ and 
$ m_{\rm n} m_{\rm p} \approx 1.7\times 10^{-2} m_0^2.$ }
\label{fig.7}
\end{figure}
\end{center}

In the weak field limit, the Hall coefficient and the carrier concentration has the form\cite{ref8}
\begin{equation}
R_{\rm H } = \frac{3 \pi}{ 8 |e| } \frac{ p-nb^2 }{ \left( p+nb\right)^2 } 
	 \approx - \frac{3 \pi}{8 |e| n} , \label{eq.11}
\end{equation}
where we define $b = \mu_{\rm n} / \mu_{\rm p}$  
and the hole parts are neglected because  $b\approx 100$ 
for InSb\cite{ref10,ref11}.  
Equation (\ref{eq.11}) and Fig.~\ref{fig.3}(b) gives the carrier concentration
of the electron in Fig.~\ref{fig.7}. 
We can fit the following function of the temperature by the least square method as 
\begin{equation}
n(T) = 3.3 \times 10^{20} T^{1.5} \exp \left( - \frac{2600}{k T}\right). \label{eq.12}
\end{equation}

We assume that the sample is in the intrinsic region near room temperature. The carrier concentration of the intrinsic semiconductor is written by\cite{ref8}
\begin{equation}
n_{\rm i} (T) = 2\left( \frac{ \sqrt{m_{\rm n} m_{\rm p} } k T}{2\pi \hbar^2 } \right)^{\frac{3}{2} } 
      \exp \left( - \frac{ E_{\rm G}}{2kT} \right), \label{eq.13}
\end{equation}
where $E_{\rm G}$ is the energy gap, $m_{\rm n}$ the effective mass of the electron, and $m_{\rm p}$ the effective mass of the hole. 
Comparing eq. (\ref{eq.12}) with eq. (\ref{eq.13}), 
we obtain 
\begin{eqnarray}
E_{\rm G} &\approx& 2600 \times \frac{k}{|e|} = 0.22 \mbox{ [eV] }, \label{eq.14} \\
m_{\rm n} m_{\rm p} &\approx& \left\{ 
       \left( \frac{3.3\times 10^{20}}{2} \right)^{2/3} 
       \frac{2\pi \hbar^2}{k}  \right\}^2 \approx 1.7\times 10^{-2} m_0^2. \label{eq.15}
\end{eqnarray}

\subsection{Mobility of electron}
\begin{center}
\begin{figure}
\epsfxsize=6cm \epsffile{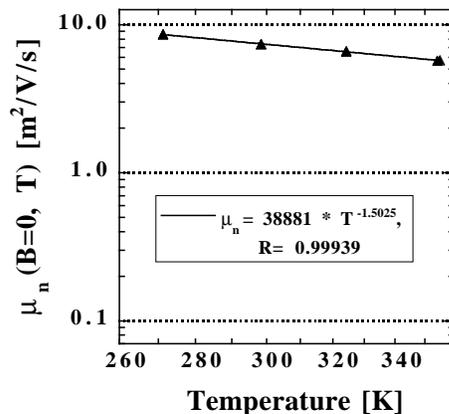}
%\epsfile{width=6cm,file=fig8.eps} 
\caption{Temperature dependencd of the electron mobility.
The experimental result are fitted by $\mu_{\rm n} \approx 7.5 \times ( T/300 )^{-1.50} $
[ m$^2$/V/s ]. }
\label{fig.8}
\end{figure}
\end{center}

In the weak field limit, the Boltzmann equation for the single parabolic non-degenerated band model gives the conductivity as
follows\cite{ref8}:
\begin{equation}
\sigma = 1/\rho = |e| n \mu_{\rm n} \left( 1+ \frac{p}{nb} \right)
  \approx |e| n \mu_{\rm n}, \label{eq.16}
\end{equation}
where we use $b \ll 1$ for InSb. Equations (\ref{eq.11}) and (\ref{eq.16})
give the mobility of the electron in Fig.~\ref{fig.8}. 
The temperature dependence of the mobility of the electron is obtained as
\begin{equation}
\mu_{\rm n} \approx 7.5 \times \left( \frac{T}{300} \right)^{-1.50}
\mbox{[m$^2$/V/s]}. \label{eq.17}
\end{equation}
The exponent -1.5 denotes that the dominant scattering process is acoustic phonon scattering.

\begin{center}
\begin{figure}
\epsfxsize=6cm \epsffile{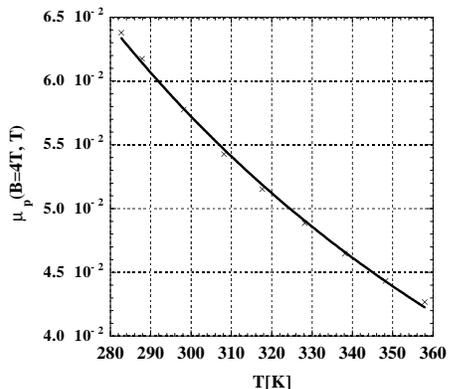}
%\epsfile{width=6cm,file=fig9.eps} 
\caption{Temperature dependencd of the hole mobility from the Nernst coefficients.
The experimental result are fitted by $\mu_{\rm p} (B=4 \mbox{Tesla}) 
\approx 0.065 \times ( T/300 )^{-1.7} $
[ m$^2$/V/s ]. }
\label{fig.9}
\end{figure}
\end{center}

\subsection{Mobility of hole}
In the strong magnetic field limit, the Nernst coefficient of the intrinsic semiconductor becomes\cite{ref10}
\begin{equation}
N \approx \frac{k}{|e|} \mu_{\rm p} \left( 4 +\frac{E_{\rm G} }{kT} \right). \label{eq.18}
\end{equation}
Substituting the values of the Nernst coefficient  given by the experiment at $B =4$ Tesla in the eq.~(\ref{eq.18}), 
we obtain the mobility of the hole in Fig.~\ref{fig.9}.
By the least square method, we also have the temperature dependence of the hole mobility as
\begin{equation}
\mu_{\rm p} (B=4\mbox{Tesla}) \approx 0.065\times 
\left( \frac{T}{300} \right)^{-1.7} \mbox{[m$^2$/V/s]} . \label{eq.19}
\end{equation}

\subsection{Fermi level}
\begin{center}
\begin{figure}
\epsfxsize=6cm \epsffile{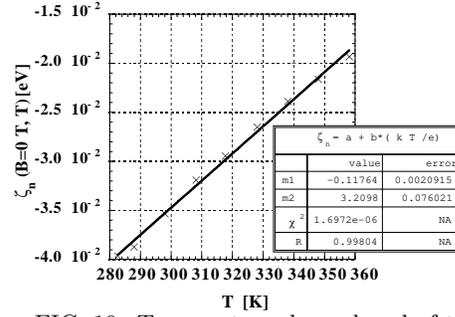}
%\epsfile{width=6cm,file=fig10.eps} 
\caption{Temperature dependencd of the fermi level.
The experimental result are fitted by $ \zeta_{\rm n} = - 0.117 + 3.2 \frac{kT}{|e|} $
[ eV ].  This equation denotes that $ E_{\rm G} \approx 0.23 $ [ eV ] and 
$ m_{\rm p}/ m_{\rm n} \approx 73.$ }
\label{fig.10}
\end{figure}
\end{center}

The thermoelectric power gives the Fermi level as follows\cite{ref8}:
\begin{eqnarray}
\alpha &=& \alpha_{\rm n} \frac{\sigma_{\rm n} }{ \sigma_{\rm n} + \sigma_{\rm p} }
 + \alpha_{\rm p} \frac{\sigma_{\rm p} }{\sigma_{\rm n} + \sigma_{\rm p} } \nonumber \\
&=& \alpha_{\rm n} \left( \frac{1}{ 1+\frac{p}{nb} } \right)
  + \alpha_{\rm p} \left( \frac{ \frac{p}{nb} }{ 1+\frac{p}{nb} } \right) \nonumber \\
&\approx& \alpha_{\rm n } = - \frac{k}{ |e| } 
   \left( 2 - \frac{ \zeta_{\rm n} }{kT} \right), \label{eq.20}
\end{eqnarray}
where is the Fermi level from the edge of the conduction band and negative. Equation~(\ref{eq.20}) and the experimental results of the thermoelectric power give the Fermi level in Fig.~\ref{fig.10}.
In the intrinsic region, the Fermi level becomes\cite{ref8} 
\begin{equation}
\zeta_{\rm n} = - \frac{E_{\rm G} }{2} + \frac{3kT}{4|e|} \ln 
\left( \frac{m_{\rm p}}{m_{\rm n} } \right) \ \ \ \mbox{[eV]}. \label{eq.21}
\end{equation}
Analysis in Fig.~\ref{fig.10} 
gives the temperature dependence of the Fermi level as follows
\begin{equation}
\zeta_{\rm n} = - 0.117 + 3.2 \frac{kT}{|e|} \ \ \ \mbox{[eV]}. \label{eq.22}
\end{equation}
From eqs.~(\ref{eq.21}) and (\ref{eq.22}), 
the energy gap and the ratio of the effective masses of the electron and the hole is given as
\begin{equation}
E_{\rm G} \approx 0.23 \ \ \  \mbox{ [ eV ] }, 
\ \ \frac{m_{\rm p} }{m_{\rm n} } \approx 73. \label{eq.24}
\end{equation}
\newpage

\section{Discussion and conclusions}
We summarize basic physical quantities obtained by the experiment in Table~\ref{tab.2}, 
where the reference values are also written. This table shows that the experimental results are almost coincident with the previous works. 
It is concluded that the Hall coefficient, the conductivity and 
the thermoelectric power of InSb near room temperature in the weak field 
are given the Boltzmann equation for the non-degenerate parabolic 
two-band with the acoustic phonon scattering. However, the Nernst coefficient 
is  very smaller than the theoretical value in the weak field. The behavior of the Nernst coefficient in the strong magnetic field is consistent 
with the two-band model. We try to explain the difference between 
the experimental and the theoretical values of the Nernst coefficient 
in the weak field. Moreover we will measure the thermal conductivity 
in the magnetic field to estimate the figure of merit.  

\begin{center}
\begin{table}
\epsfxsize=8cm \epsffile{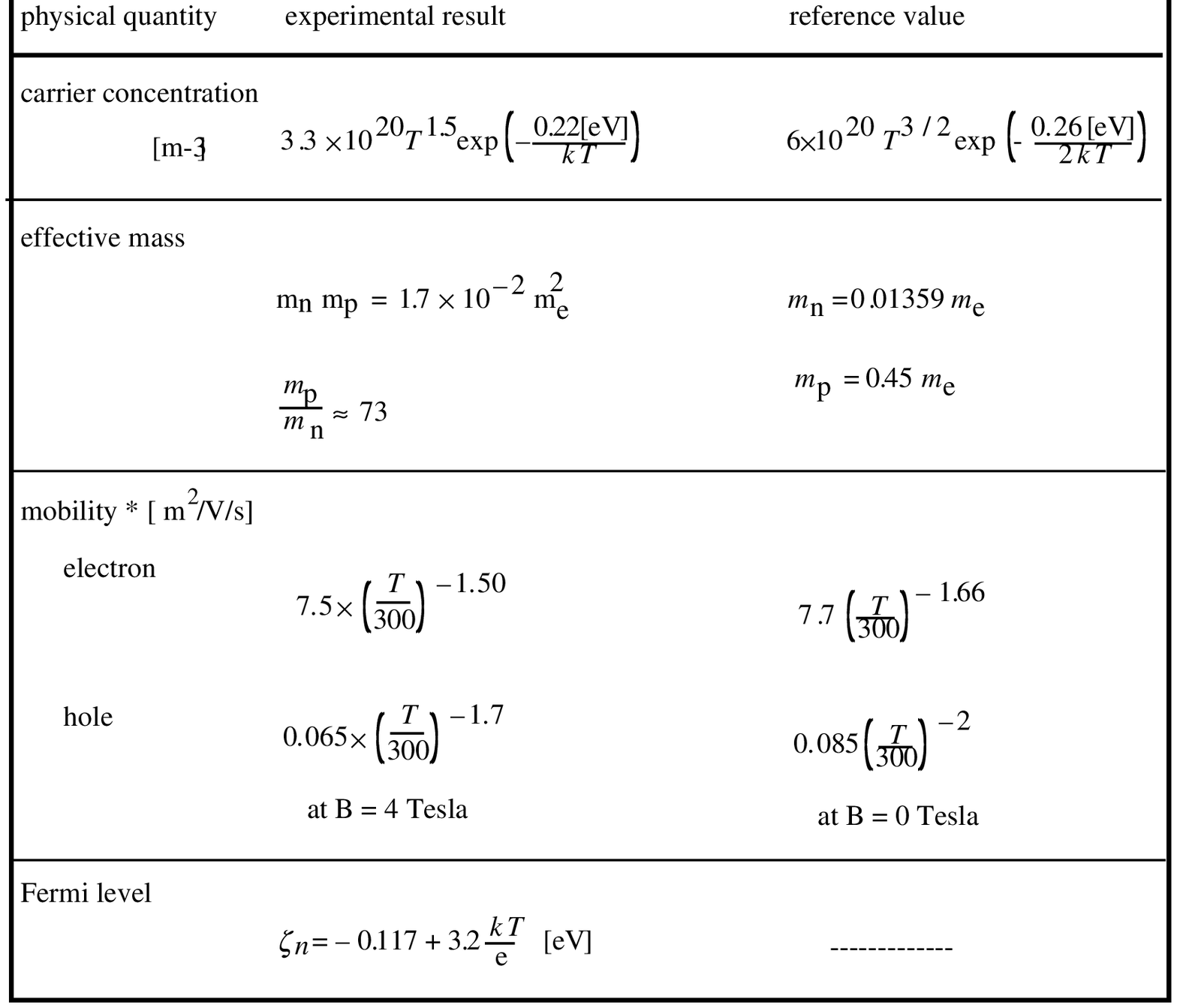}
%\epsfile{width=8cm,file=tab2.eps} 
\caption{Comparision of the experimental results and 
the theoretical vales.}
\label{tab.2}
\end{table}
\end{center}

\acknowledgments
The authors are grateful to Dr. Tatsumi in Sumitomo Electric Industries 
and Prof. Kuroda in Nagoya university for providing semiconductors.
We appreciate Prof. Iiyoshi in the National 
Institute for Fusion Science 
for his helpful comments.

\end{document}